\documentclass[twocolumn,numbering,showpacs]{revtex4-1}

\usepackage{graphicx,color}
\usepackage{latexsym}
\usepackage{amsmath,amssymb}        
\usepackage[draft=false]{hyperref}
\usepackage{mathrsfs}
\usepackage{comment}
\usepackage{soul}
\usepackage{mathrsfs}
\definecolor{purple}{rgb}{0.58,0.0,0.83}
\definecolor{orange}{rgb}{1,0.5,0}
\DeclareSymbolFontAlphabet{\mathrsfs}{rsfs}
\DeclareMathAlphabet{\mathcal}{OMS}{cmsy}{m}{n}

\begin{document}


\title{Oscillation modes of ultralight BEC dark matter cores}


\author{F. S. Guzm\'an}
\affiliation{Laboratorio de Inteligencia Artificial y Superc\'omputo,
	      Instituto de F\'{\i}sica y Matem\'{a}ticas, Universidad
              Michoacana de San Nicol\'as de Hidalgo. Edificio C-3, Cd.
              Universitaria, 58040 Morelia, Michoac\'{a}n,
              M\'{e}xico.}


\date{\today}


\begin{abstract}
Structure formation simulations of ultralight bosonic dark matter, ruled by the Gross-Pitaevskii-Poisson (GPP) system of equations, show that dark matter clumps to form structures with density profiles consisting of a core surrounded by a power-law distribution. The core has a density profile similar to that of spherically symmetric equilibrium configurations of the GPP system. These configurations have been shown to be stable under a variety of perturbations and  to have  attractor properties. It is interesting to know the dominant frequencies of oscillation of these configurations. One reason is that in galaxies the effects of perturbations can trigger observable effects for specific frequency modes. Another reason is that during the process of structure formation, the oscillation modes can help to characterize a core. Based on these motivations, in this manuscript we present a systematic numerical analysis of the reaction of equilibrium configurations to various axi-symmetric perturbations with modes $l=0,1,2,3,4$. We then calculate the first few oscillation frequencies of equilibrium configurations for each mode. 
\end{abstract}


\pacs{keywords: dark matter -- Bose condensates}


\maketitle

\section{Introduction}
\label{sec:introduction}

Ultralight spin-less dark matter is one of the dark matter candidates that shows properties that suit the current observational restrictions. Among the convenient properties of this particle are that it is ultralight, with mass of order $m\sim 10^{-22}eV$ \cite{Matos-Urena:2000,Hlozek:2014,Ostriker:2016}, which implies it has a de Broglie wave-length that can prevent structures from developing cusps at their centers, which may be a solution of the cusp-core problem \citep{Chen:2016,Schive:2015,Du:2016,Robles:2013,Bernal:2016}. At larger scales, there are now structure formation simulations showing that this candidate behaves just like CDM \cite{Schive:2014,Schive:2014hza}. Another problem possibly addressed by this candidate is the abundance of substructure  \cite{Matos-Urena:2000,Ostriker:2016,Marsh-Ferreira:2010,Hlozek:2014,Schive:2015}. An appealing property is that for bosons with such a small mass, the critical condensation temperature is high, of order  $T_c\sim 1/m^{5/3}$ in the TeV scale, between the electroweak and quark epochs, at very early stages between $t \sim 10^{-32}$ and $10^{-12}$s after the Big Bang \cite{JuanMagana}.

However, on local scales some interesting problems remain, such as the relaxation processes of structures that can involve the gravitational cooling process \cite{GuzmanUrena2003,GuzmanUrena2006}, dynamical friction \cite{Mocz2017} or appropriate damping terms  \cite{ChavanisDamping}. Nevertheless, structure formation simulations agree in that ultralight bosonic matter clumps into structures with universal profiles composed of a core, sometimes called solitonic profile, and a surrounding cloud of dark matter that suits a NFW profile \cite{Schive:2014,Schive:2014hza,Schwabe:2016,Mocz2017,ChavanisCore}.

Various oscillations of cored structures have been found to be interesting and potentially important, in particular the quasinormal modes of equilibrium configurations presented in \cite{GuzmanUrena2006}, which are associated to the oscillation of cores formed in structure formation simulations \cite{NewRef}. The implications of these effects have been already translated into potentially observable effects on star clusters \cite{Marsh2018}. In a different context, the configuration resulting from head-on mergers has oscillations where the density changes orders of magnitude, an effect that eventually could affect the survival of luminous structures living inside such dynamical environment \cite{AvilopGuzman2018b}. For instance, in \cite{GonzalezGuzman2016} it was shown that luminous matter, modeled as as an $N-$-body system within the gravitational potential of the condensate after a merger, might not survive such violent processes of relaxation and would  be spelled out due to violent oscillations of the final configuration. In a less cataclysmic scenario,  during the structure formation process, a core is constantly being perturbed by structures moving around it \cite{Mocz2017}.  If some of the oscillations can reveal whether the model is viable or provide the fingerprints of cores, it is important to analyze how the latter respond to different types of perturbations. Oscillations in a way, may impose important restrictions or confirm the viability of the model, which is a strong a motivation to present a systematic analysis of the response of cores to perturbations in order to know the essential response modes that can eventually be identified with observations or predictions. Equally important is the construction of the oscillations spectrum of cores, which is important for their characterization. These are some reasons to carry out the present analysis, which focuses on the response of cores to various types of perturbations.

For this analysis we have two basic assumptions, the first one is that the dynamics of ultralight BEC dark matter structures is ruled by the GPP system of equations, which is the combination of the Gross-Pitaevskii equation for the condensate \citep{Pitaevskii:2003}, and Poisson's equation sourced by the density of the condensate itself, whose solution provides the potential trap that contains the condensate bounded in a region. The second assumption is that cores are spherically symmetric equilibrium configurations of the GPP system \cite{GuzmanUrena2004,GuzmanUrena2006} as commonly assumed  \cite{Schive:2014,Schive:2014hza,Schwabe:2016,Mocz2017}.

In addition, one can take a perturbative approach since it is well known these equilibrium configurations are stable under a variety of spherical and axial perturbations, and they also show a late-time attractor behavior, which means that eventually fluctuations with an arbitrary initial profile would tend toward one of these equilibrium configurations \cite{GuzmanUrena2003,GuzmanUrena2006,BernalGuzman2006a}.
Therefore, the method used here to study the modes triggered by various perturbations, consists in the  application of perturbations to equilibrium configurations, solve the full GPP system of equations in time and diagnose the evolution of physical quantities of the system. In order to be concise, in this paper we concentrate the analysis to axially symmetric perturbations only, and on the resulting oscillation modes that are excited.

The paper is organized as follows. In Section \ref{sec:methods} we descrbe the perturbations used and the analysis in the frequency domain. In Section \ref{sec:results} the frequency modes are presented and in Section \ref{sec:final} we draw some final comments and observations.

\section{Perturbations and numerical set up}
\label{sec:methods}

The GPP system of equations written in variables absorbing the constants $\hbar$, $G$, the boson mass $m$  and the $s$-wave scattering length between bosons is

\begin{eqnarray}
i\frac{\partial \Psi}{\partial t} &=& -\frac{1}{2}\nabla^2 \Psi + V\Psi + \Lambda |\Psi|^2 \Psi \nonumber\\
\nabla^2 V &=& 4\pi |\Psi|^2
\label{eq:gpp}
\end{eqnarray}

\noindent where $\Psi=\Psi(t,{\bf x})$ is the parameter order describing the macroscopic behavior of the boson gas, $\rho=|\Psi|^2$ is the gas density, $\Lambda$ is the self-interaction coefficient, and the potential $V=V(t,{\bf x})$ is the gravitational potential sourced by the gas density itself, which in general can depend on time if $|\Psi|^2$ does. Here, as in most of the recent analysis of structure formation and interaction between structures, we assume $\Lambda=0$, which is the free field case commonly named the fuzzy dark matter regime \cite{Hu2000}.

\subsection{Perturbing equilibrium configurations}

These configurations are constructed assuming a harmonic time dependence of the wave function $\Psi=e^{-i\gamma t}\psi({\bf x})$ and spherical symmetry $\psi=\psi(r)$, where the label $e$ stands for equilibrium. In this case the equations above reduce to a stationary system because $|\Psi|^2$ is time-independent, which implies $V= V(r)$ is also-time independent and Schr\"odinger's equation can be cast as a stationary equation with solution  $\psi_e=\psi_e(r)$. The resulting equations are solved in the domain $r\in [0,r_{max}]$ as a Sturm-Liouville problem for the eigenvalue  $\gamma$ as described in \cite{GuzmanUrena2004}, with boundary conditions of regularity at the origin for $\psi_e(r)$ and $V(r)$,  isolation conditions  at the outer boundary,  namely a monopolar potential satisfying $-M/r_{max}$ and $\psi_e$, together with the requirement that its radial derivative should vanish within a given tolerance at $r_{max}$.

The solution defines a one-parameter family of solutions that relates the central value of the wave function $\psi_e(0)$ and the eigenfrequency $\gamma$. Due to the known scale transform that leaves the system (\ref{eq:gpp}) invariant, $\{ t,x,\Psi \}\rightarrow \{\hat{t}/\lambda^2,\hat{x}/\lambda, \lambda^2\hat{\Psi}\}$ for an arbitrary value of $\lambda$, if we solve for one central value $\psi_e(0)$ then the whole family for all other possible values of this central value will be automatically found. The workhorse configuration corresponds to the case $\psi(0)=1$, which in turn implies that the central density is also 1, has an eigenvalue $\gamma=0.69223$,  mass $M=2.0622$ and $r_{95}=3.93$ \cite{GuzmanUrena2004}, all of which  will be helpful later on. The scale invariance allows one to study the properties, evolution, and -for the purpose of this paper- the reaction to perturbations of only this workhorse configuration, and the results can be rescaled to all the other possible configurations. This is the reason why we will perturb this particular configuration in the reminder of the manuscript.

\subsection{Perturbations}

For our analysis, the perturbation  applied to the wave function of the equilibrium configuration is a spherical shell with thickness $\sigma_p$, centered at a distance $r_p$ from the center of the configuration itself, launched with radial wave number $k_r$ and shape proportional to specific spherical harmonics. Since we only consider axially symmetric perturbations, we will only involve modes $(l,m)$ with $m=0$ and study the cases $l=0,1,2,3,4$. The perturbation profile is inspired by that in \cite{BernalGuzman2006b} where it was used to show the stability of equilibrium configurations against non-spherical perturbations, and implemented also in \cite{AvilopGuzman2018b} to understand an oscillation mode remaining after the head-on merger of two ultralight bosonic galactic cores. The perturbation profile has the following explicit form

\begin{eqnarray}
\Psi_0 &=& \psi_e + \delta \Psi, \nonumber\\
\delta\Psi &=& e^{-ik_r \sqrt{\varrho^2+z^2}}a_{l0}Y^{0}_{l},\nonumber\\
a_{l0} &=& A_{l0}e^{-(\sqrt{\varrho^2+z^2}-r_p)^2/\sigma_{p}^2} .
\label{eq:initialwfnonspherical}
\end{eqnarray}

\noindent where $\psi_e$ is the wave function of the equilibrium configuration. In order to have a consistent gravitational potential, one solves Poisson equation sourced by the density $|\Psi_0|^2$ at initial time. Then we track the evolution of the perturbed system  by numerically solving (\ref{eq:gpp}) and analyze its behavior.

We use perturbations proportional to each mode $l$ separately, in order to study how each mode affects the configuration. For this we ran a number of simulations with various values of the parameters in (\ref{eq:initialwfnonspherical}) with the aim of determining peaks in the frequency domain that are perturbation independent. Thus, for each value of $l$ we use shells  centered at two different radii $r_p=2r_{95},~4r_{95}$, two different thickness of the shell $\sigma_p=1,~2$ and three different wave number values $k_r=0,1,2$. For each combination of these parameters, the amplitudes $A_{l0}$ were fine-tuned in such a way that the initial total mass of the perturbed configuration is $M=M_e+\delta M$, where $M_e=2.0622$ is the mass of the equilibrium configuration in standard code units (as in \cite{GuzmanUrena2004}). Four values of $\delta M$ were used $\delta M=0.1\%,~0.5\%, ~1\%$ and $1.5\%$ of $M_e$, where $\delta M>0$, meaning that the perturbation adds matter to the equilibrium configuration.

\subsection{The code}

In order to follow the evolution of the perturbed configuration, we solve the GPP system (\ref{eq:gpp}) with initial conditions given by  (\ref{eq:initialwfnonspherical}). For this we use an upgraded version of the code introduced in \cite{BernalGuzman2006a,BernalGuzman2006b}; this code uses cylindrical coordinates to solve the equations in the space-time domain $\varrho\in[0,\varrho_{max}]\times z \in [-z_{max},z_{max}]$ for $t\ge 0$ and  two values of $\varrho_{max}=z_{max}$, specifically 20 and 40. These two domains suffice to contain multiples of the scale $r_{95}$ and more importantly, the use of two considerably different domain sizes with consistent results, as shown below, indicates the results are unaffected by boundary effects. The numerical implementation uses a finite differences method to discretize the system of equations on a two-dimensional uniformly discrete spatial domain, with resolution in a convergent regime. For the time integration we  use the method of lines with a third order accurate integrator. More details about the numerical methods can be found in \cite{BernalGuzman2006a,BernalGuzman2006b}.

\subsection{Boundary conditions} 

Boundary conditions are very important since we want to analyze oscillation modes because the wave function could be reflected from the boundaries and promote the formation of certain unphysical modes. In order to avoid this, for the wave function we impose a sponge consisting of an imaginary potential, which is non-zero  near the boundary and has the profile defined in Refs. \cite{GuzmanUrena2004,BernalGuzman2006b}. 
For the gravitational potential a multipolar expansion of the potential is used. 

\subsection {Detectors} 

It is very comfortable to analyze the evolution of the perturbed configuration by monitoring a scalar function of the density. For spherically symmetric scenarios $(l=0)$, its value at the coordinate origin is useful. Nevertheless, as we expect that the evolution can be more general in the case of non-spherical perturbations, we measure the value of the density at  Detector 1  located at $D_1=(\varrho_D,0)$ and at Detector 2 at $D_2=(0,z_D)$ with the aim of measuring the oscillation modes along the two coordinate axes. Although we use detectors at various positions along the $\varrho$ and $z$ axes, there are no differences between the frequency modes, and we only present the results for the specific location with $\varrho_D = z_D=2\simeq r_{95}/2$.


\subsection{Runs set up, output and analysis} 

In order to carry out a consistent analysis of the runs, the numerical parameters were standardized. The time domain for the runs is $t\in[0,3000]$ in code units, even though in some of the plots below a subset of this time domain is used. The production resolution is $\Delta \varrho=\Delta z=0.01\sim0.025r_{95}$ and we use the same time step $\Delta t$ in all cases. The  Courant-type factor $\Delta t / \Delta \varrho^2 = 0.2$, which implies $\Delta t=2\times 10^{-5}$. The output from the simulations are the time-series of the density at the coordinate origin and at $D_1$ and $D_2$. What is done next is to calculate the Fourier Transform (FT) of the time series with output in the frequency domain. In order for the FT to be accurate in the frequency domain, it sufficed the output of the time series to have a period of  $1000\Delta t$. For a consistency check the FT was also calculated using a downsampling time-series of the output every $2000\Delta t$.

\section{Analysis and results}
\label{sec:results}


\subsection{Spherical perturbations and consistency check}

\begin{figure}
\centering
\includegraphics[width= 4.25cm]{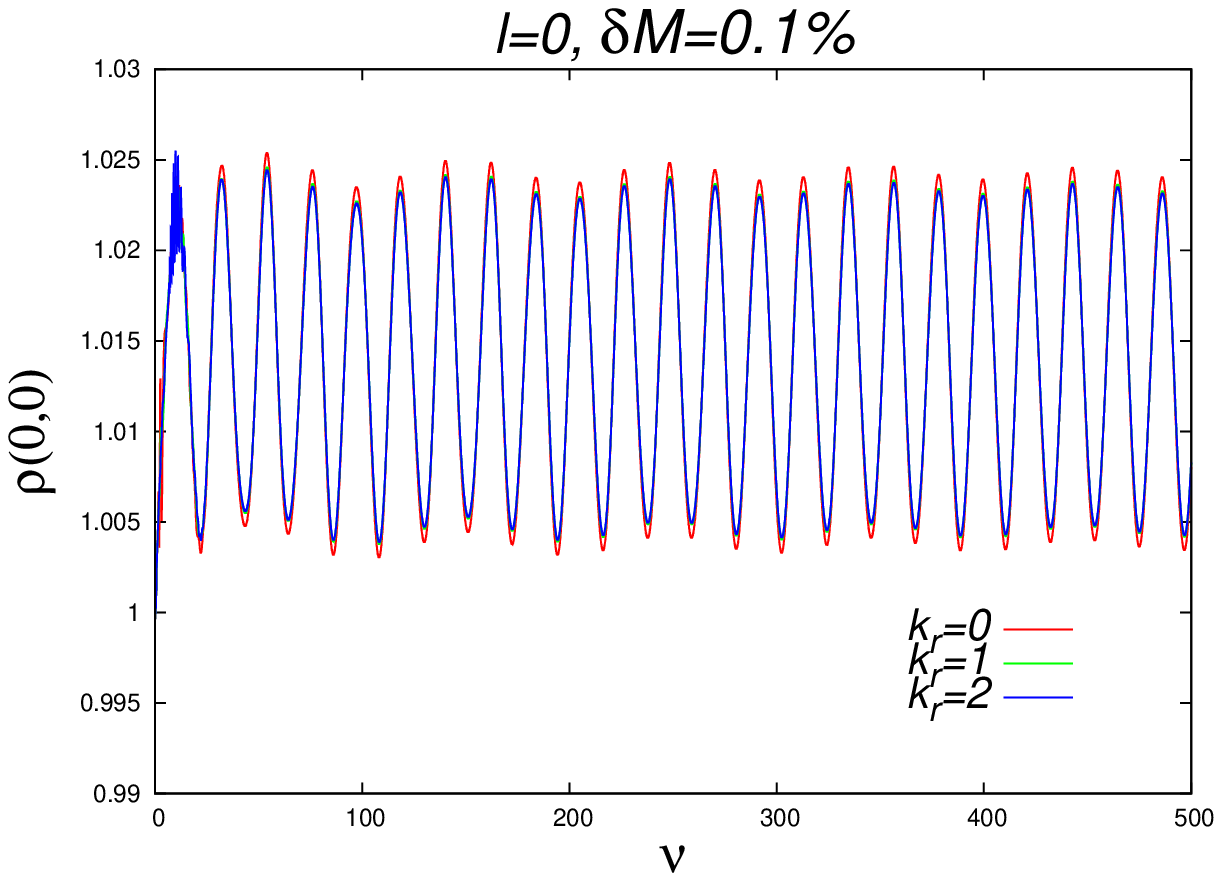}
\includegraphics[width= 4.25cm]{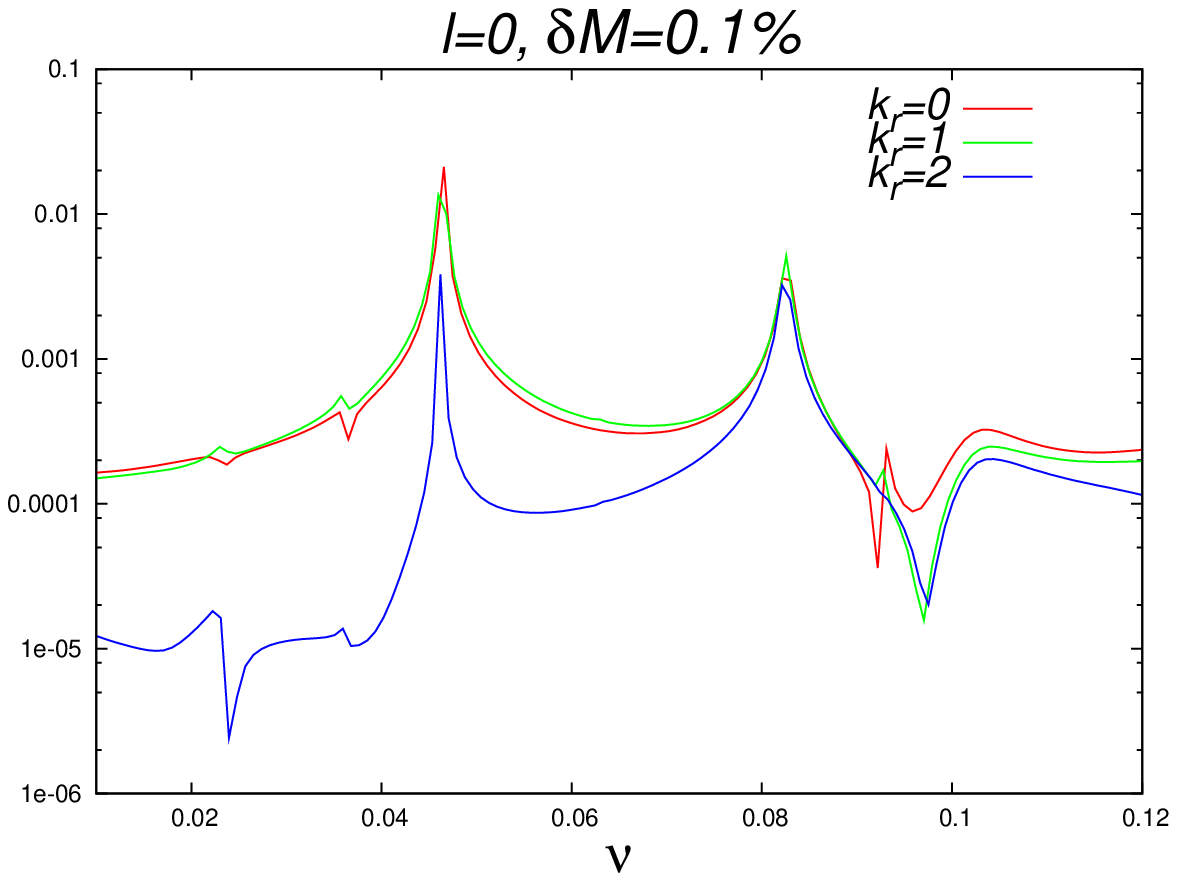}
\includegraphics[width= 4.25cm]{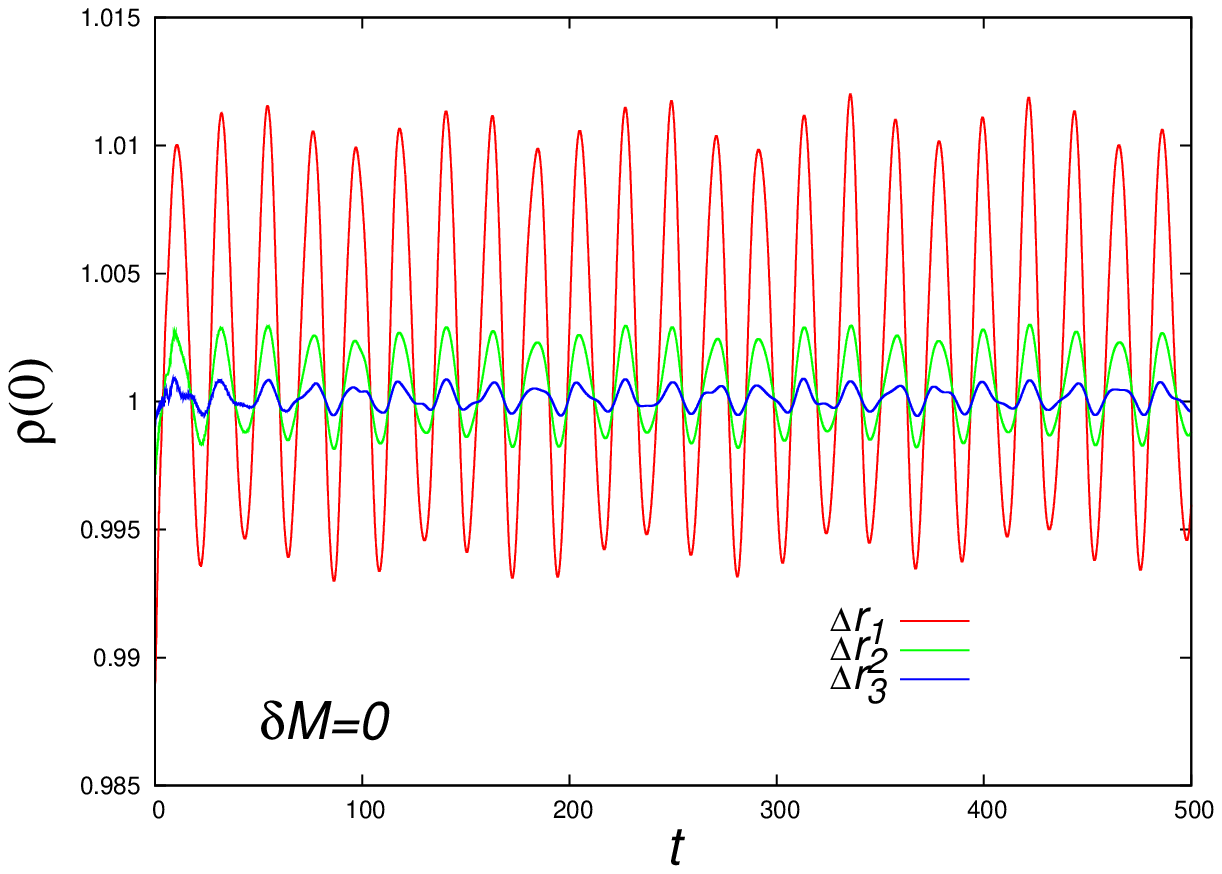}
\includegraphics[width= 4.25cm]{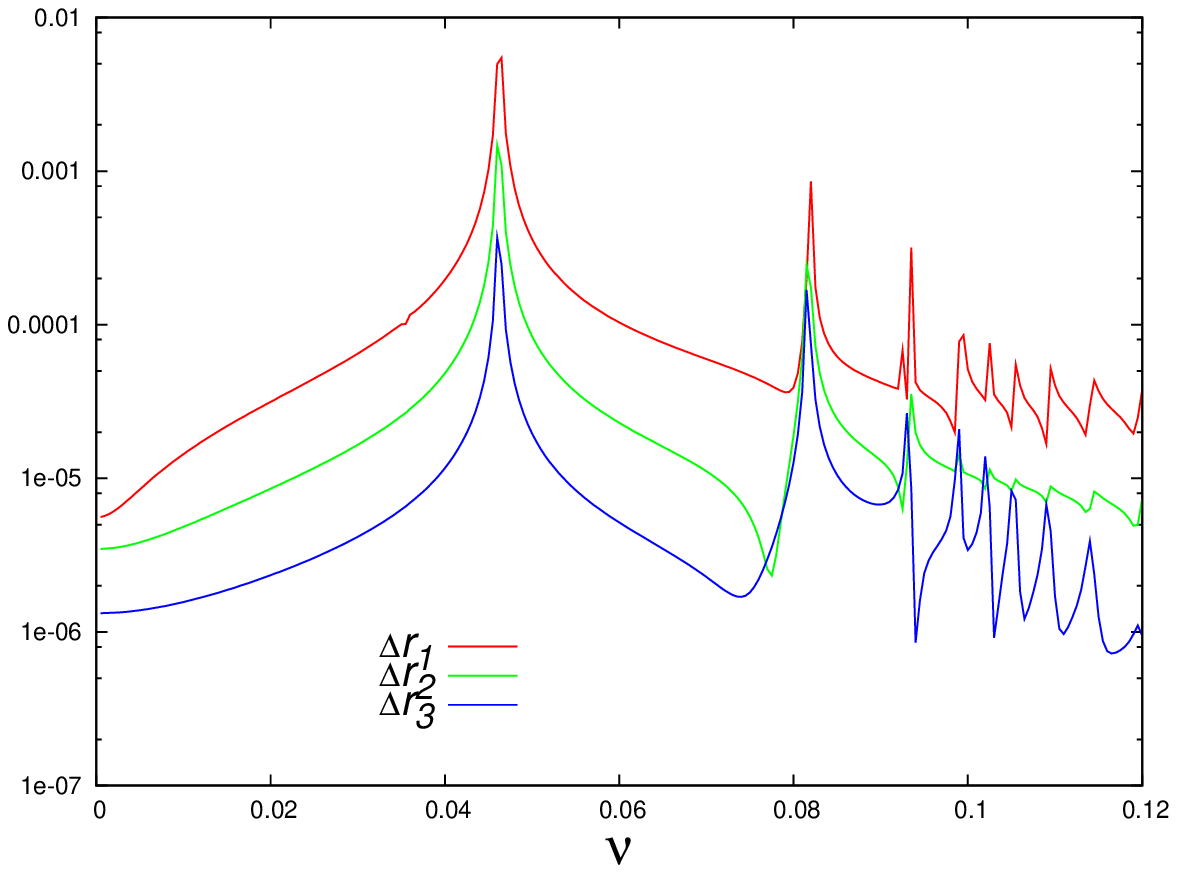}
\caption{Central value of the density in a a subset of the time domain and its FT. At the top we show the results for the case with perturbation parameters $k_r=0,1,2$, $\sigma_p=1$, $\varrho_{max}=20$ and $\delta M = 0.1\%$. As a consistency check, in the bottom  we show the time series of the central value of the density calculated using a spherically symmetric code with three very high resolutions $\Delta r_3=\Delta r_2/2$, $\Delta r_1/4 = 2.5\times 10^{-3}r_{95}$, where the density oscillations are triggered by the numerical truncation errors only.}
\label{fig:numericalperturbation}
\end{figure}

We  apply a spherical perturbation using (\ref{eq:initialwfnonspherical}) for the case $l=0$, evolve the initial wave function $\Psi_0$, measure the density at detectors $D_1$, $D_2$ and calculate the FT of the resulting time series. These time-series measured by the detectors for a given time-window  and their FT appear in Figure \ref{fig:numericalperturbation} for the particular case of $\sigma_p=1$, $\varrho_{max}=20$ and $\delta M= 0.1\%$. 
As a consistency check, for this particular mode, we used an independent code that solves the GPP system in spherical symmetry, with only numerical perturbations as done in \cite{GuzmanUrena2006}, measured the density at the center as a function of time and calculated its FT. The purpose of this exercise is to verify that the peaks in the present analysis are reproduced with an independent code at least for spherical symmetry.
The result is that there are two  peaks indicating two dominating modes with frequencies at $\nu_1 \sim 0.0462$ and $\nu_2 \sim 0.0817$. The first of these frequencies was obtained using a perturbation theory analysis in \cite{GuzmanUrena2004}, which we will call fundamental mode, and the second one was numerically  in \cite{BernalGuzman2006b} within the context of the stability analysis.

\subsection{Perturbations with $l=1$}

\begin{figure}
\centering
\includegraphics[width= 4.25cm]{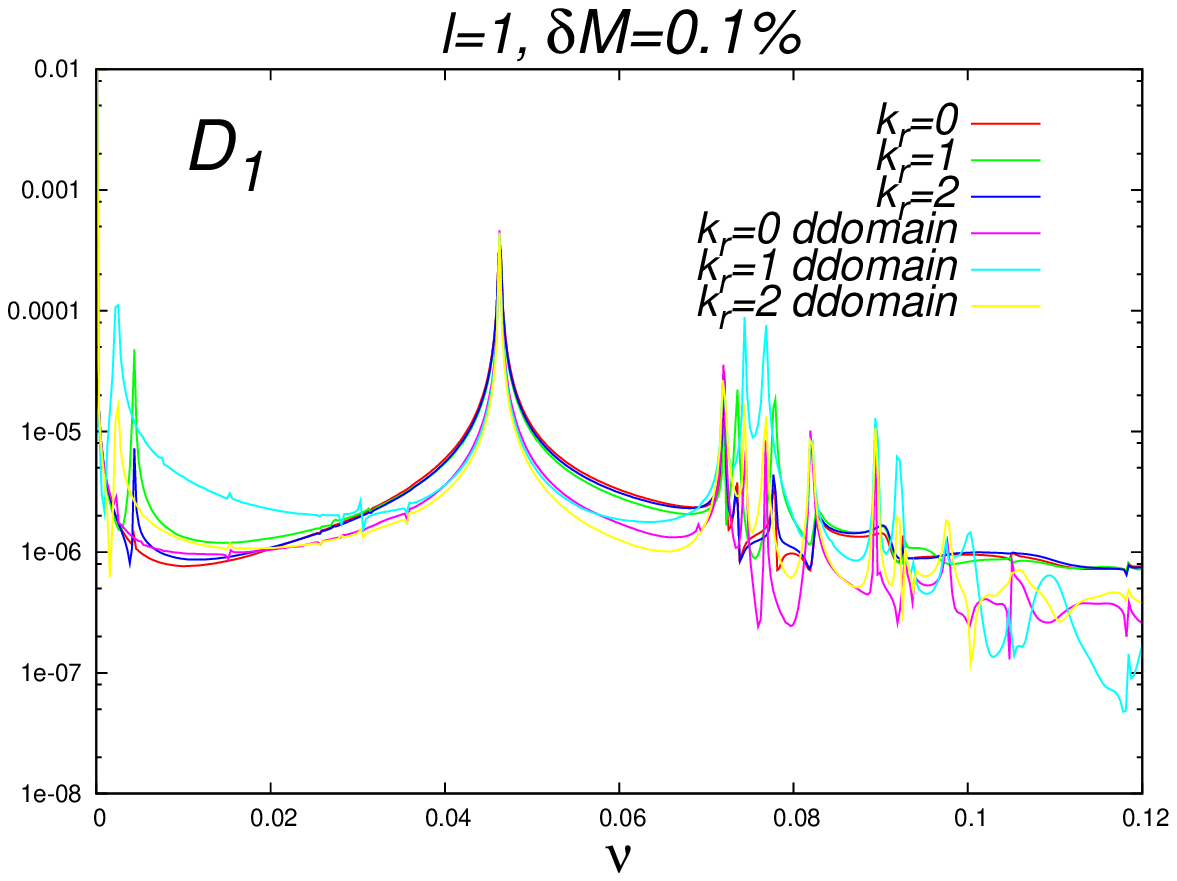}
\includegraphics[width= 4.25cm]{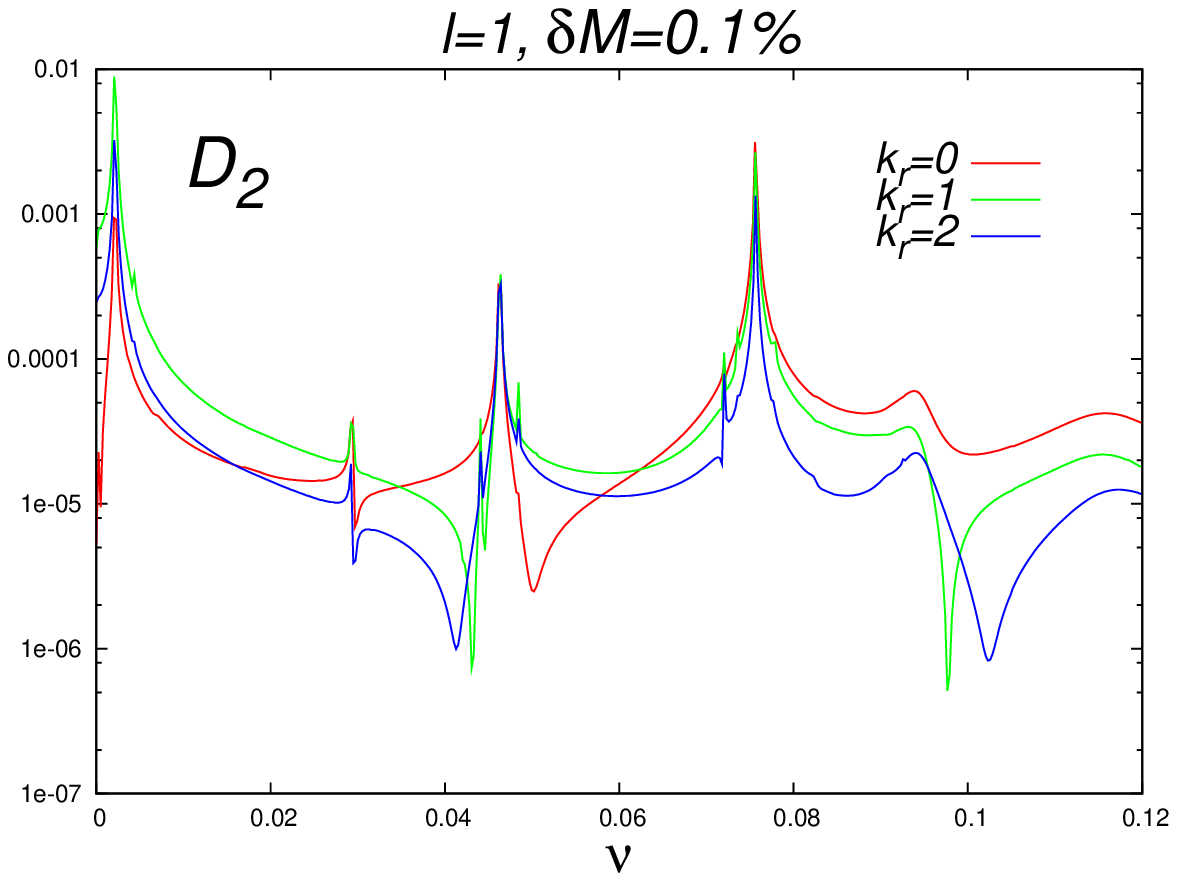}
\includegraphics[width= 4.25cm]{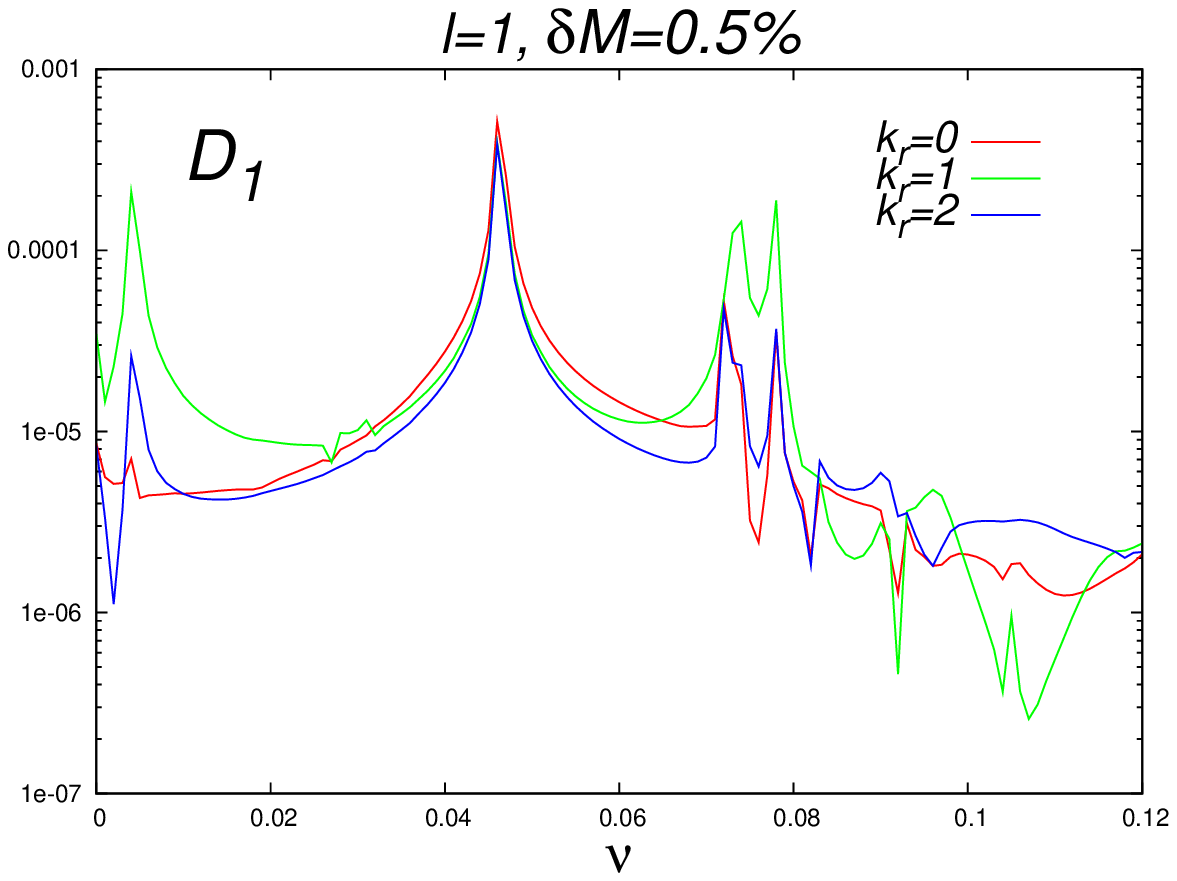}
\includegraphics[width= 4.25cm]{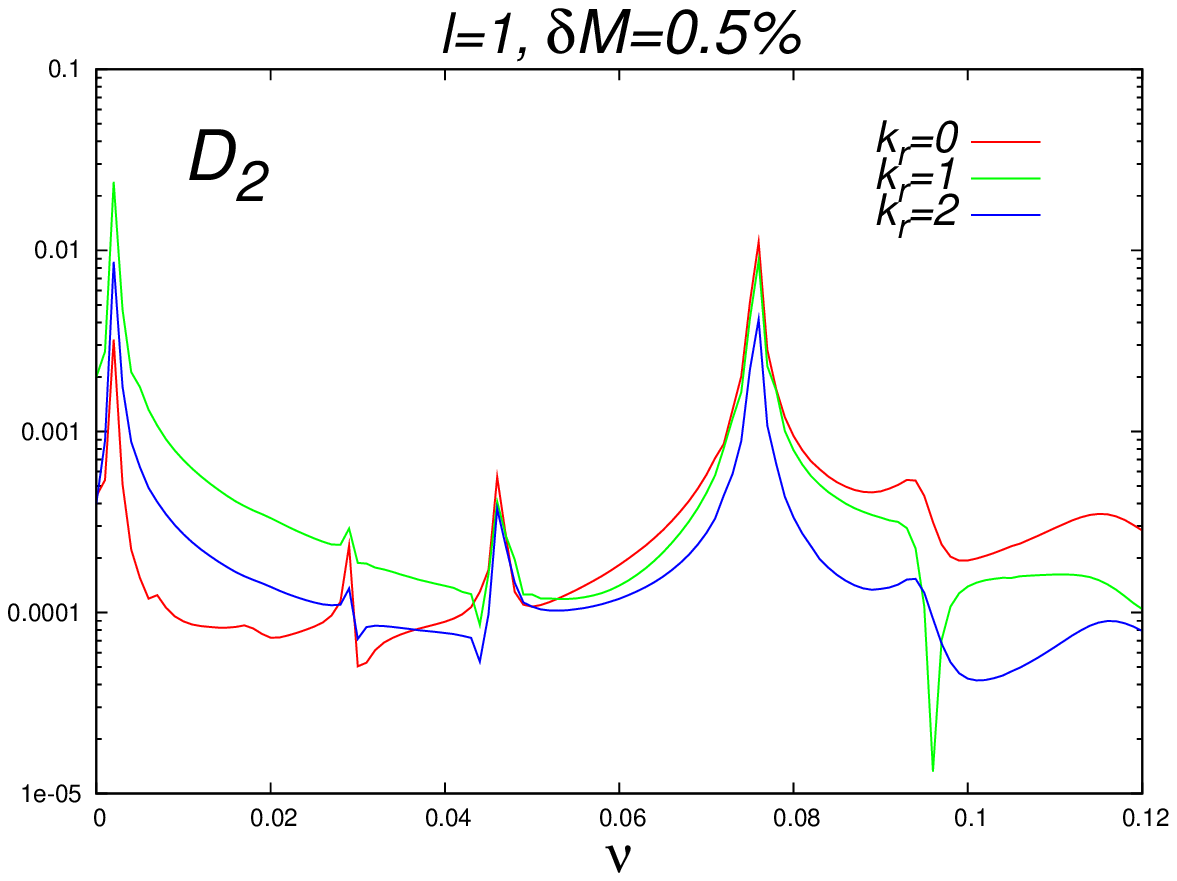}
\caption{Spectrum of the perturbations with the $l=1$ mode in the frequency domain. Shown are the results measured by  $D_1$  located at the $\varrho-$axis (left) and $D_2$ located on the $z-$axis (right).}
\label{fig:mode10}
\end{figure}

In this case we follow the prescription (\ref{eq:initialwfnonspherical}) with $l=1$. The initial $\Psi_0$ is evolved using the three values of the wave number $k_r$, the four mass contributions of the perturbation $\delta M$, the two values of $\sigma_p$ and the two numerical domains described above. We produce the time-series corresponding to the density measured at detectors $D_1$ and $D_2$. Then we calculate the FT that we show in Fig. \ref{fig:mode10}, for $\sigma_p=1$, $\rho_{max}=20$, the three values of the wave number $k_r$ and two out of the four mass contributions of the perturbation $\delta M=0.1\%,0.5\%$.

In the analysis, the frequencies we consider associated to the configuration and not to a particular feature of the perturbation, are those that show a peak for all the combinations of $\sigma_p$, $\varrho_{max}$, $k_r$ and $\delta M$, otherwise we consider it only a feature. Features are for instance the first peak measured by $D_1$ along the $\varrho-$axis with a frequency about $\nu\sim0.004$, which is not a peak for $k_r=0$ and $k_r=2$. Another one is that measured by $D_2$ with frequency $\nu\sim 0.027$, because it is not a peak for any of the values of $k_r$ and the two values of $\delta M$ shown in Fig. \ref{fig:mode10}.

\begin{figure}
\centering
\includegraphics[width= 4.25cm]{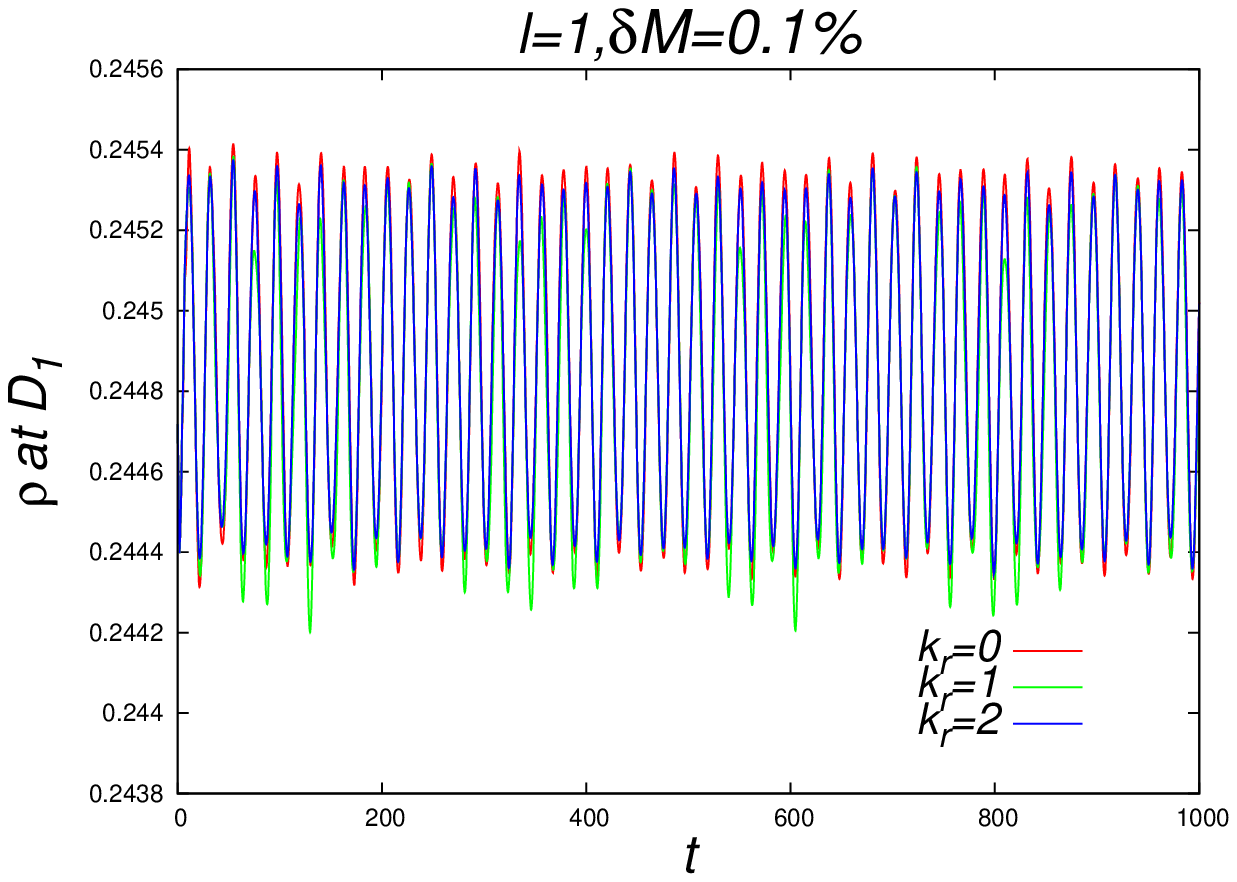}
\includegraphics[width= 4.25cm]{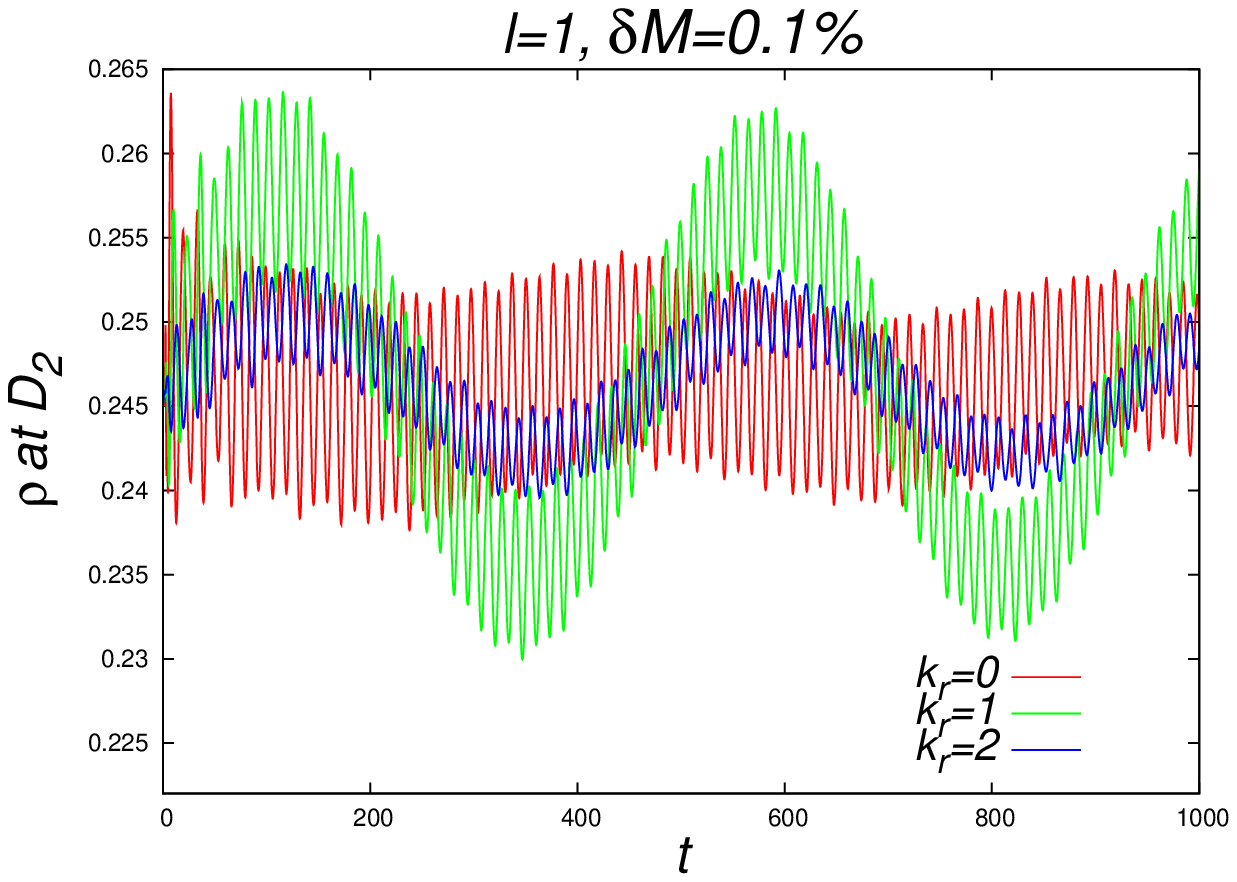}
\i\caption{Time series of the density at detector $D_1$ at the left and $D_2$ at the right, for the case $l=1$, $\sigma_p=1$, $\varrho_{max}=20$ and $\delta M = 0.1\%$. There is a low frequency mode measured clearly by $D_2$.}
\label{fig:mode10_bis}
\end{figure}

With this in mind, we now turn to discuss the peaks we consider to be perturbation independent. The first important mode triggered by the $l=1$ perturbation, measured by detectors $D_1$ and $D_2$ with respective frequencies ${}^{10}\nu^{\varrho}_{1}={}^{10}\nu^{z}_{1}=0.0462$,   corresponding to the fundamental mode. There are two other peaks measured by $D_2$: one with frequency ${}^{10}\nu^{z}_2\sim0.0756$ that is triggered only by this perturbation and one is a peak with very low frequency ${}^{10}\nu^{\varrho}_0\sim 0.002$, that deserves some comments. This mode can be seen clearly  from the time series of the density measured by detector $D_2$, as shown in Fig. \ref{fig:mode10_bis} for the perturbation with $\sigma_p=1$, $\varrho_{max}=20$ and $\delta M=0.1\%$. Its period is of around 500 units of time, that corresponds to the first peak close to the axis in the right side of Fig. \ref{fig:mode10}. This frequency mode is particularly interesting because it has been triggered only by $l=1$ and $l=3$ perturbations, and has been calculated in \cite{AvilopGuzman2018b} with the aim of explaining long period modes excited during head-on core mergers.

\begin{figure}
\centering
\includegraphics[width= 4.25cm]{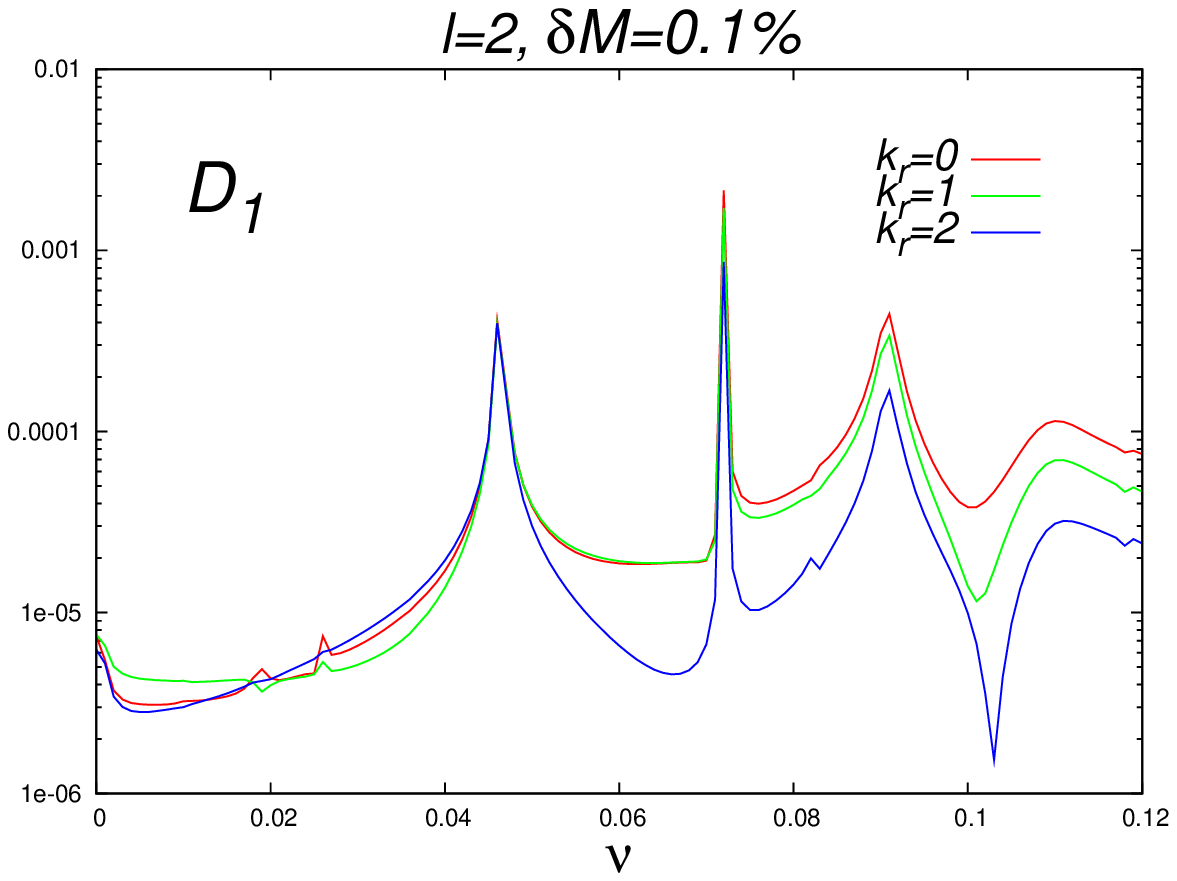}
\includegraphics[width= 4.25cm]{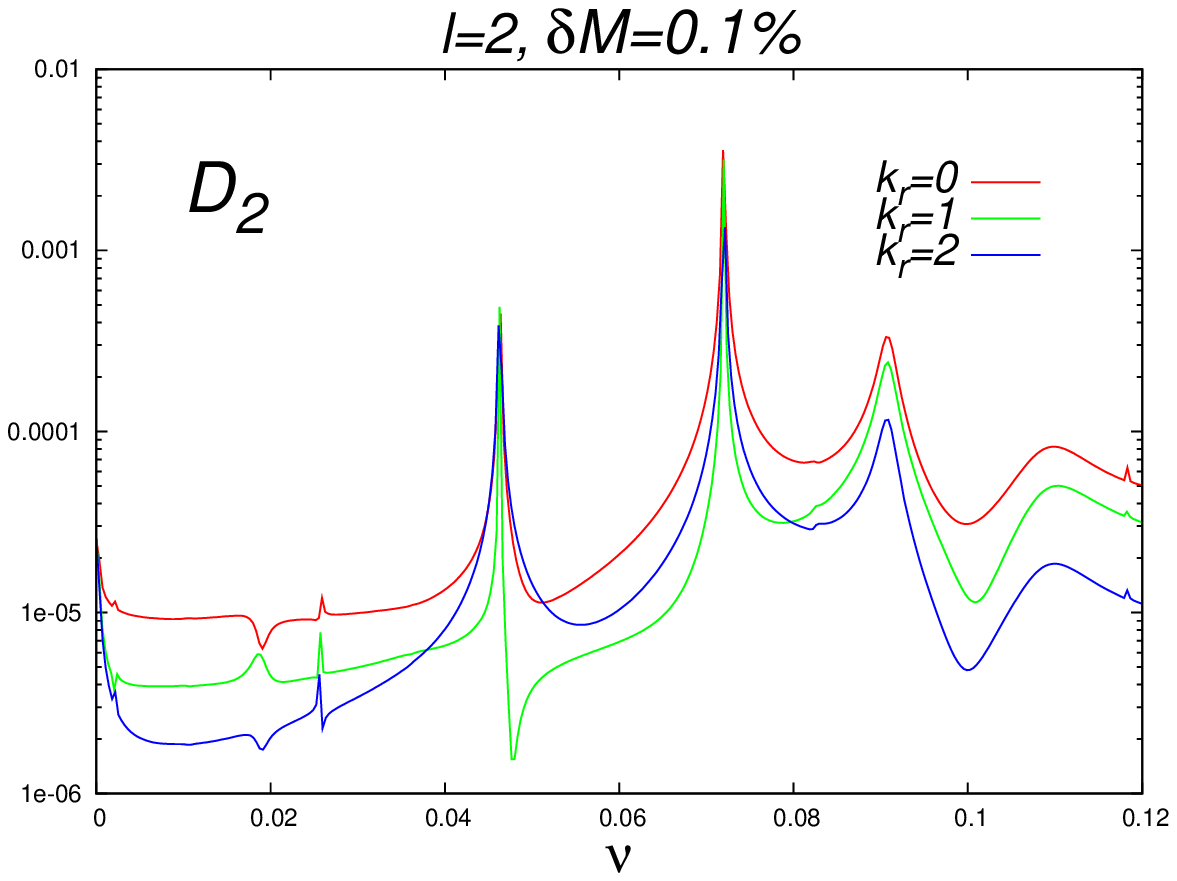}
\includegraphics[width= 4.25cm]{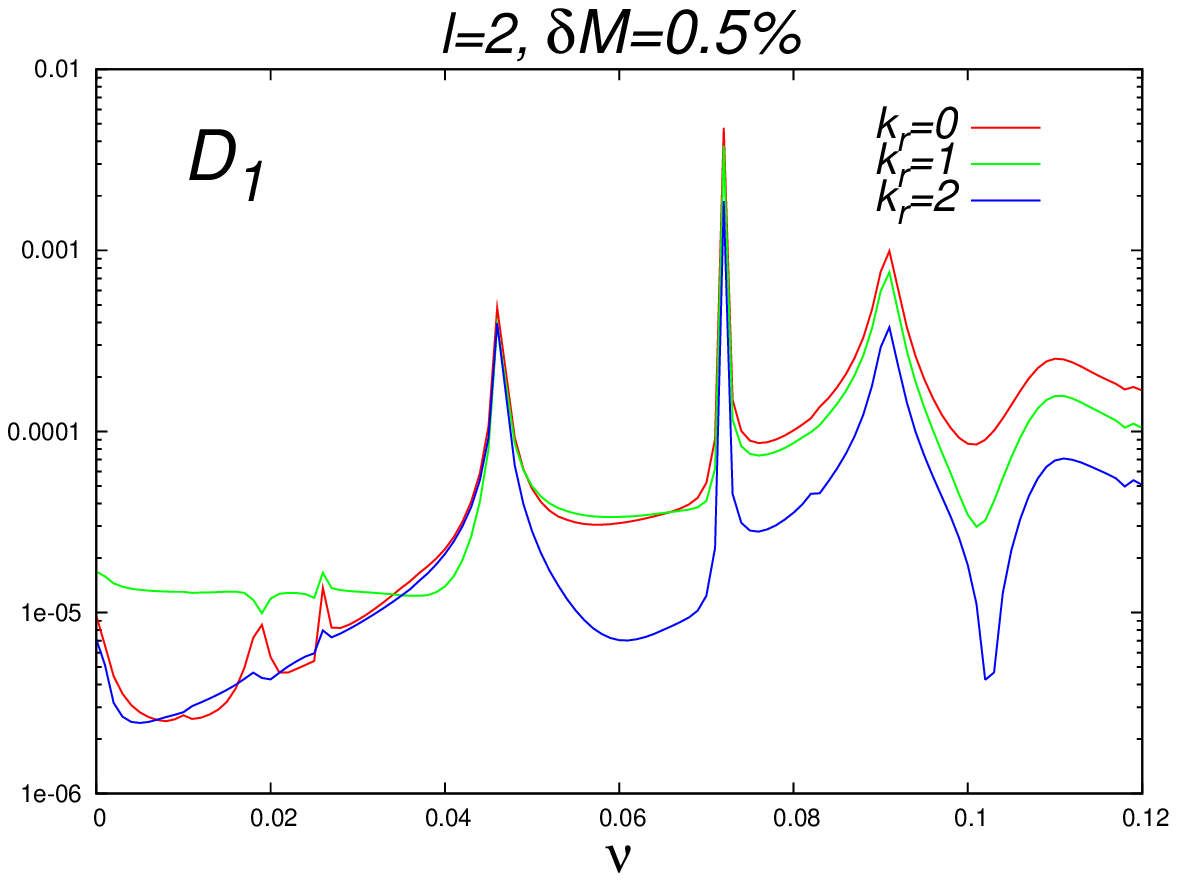}
\includegraphics[width= 4.25cm]{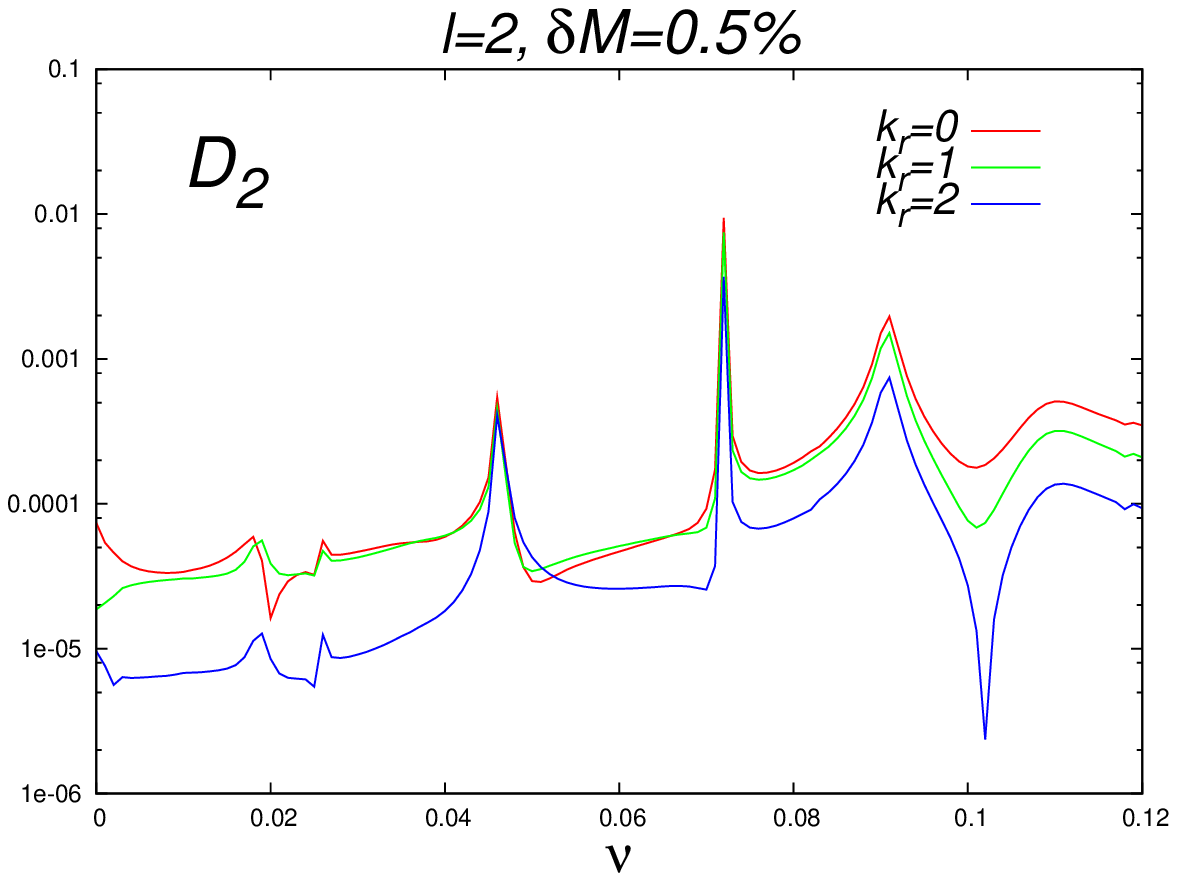}
\caption{Spectrum of the perturbations for the $l=2$ mode with $\sigma_p=1$, $\rho_{max}=20$, the three values of the wave number $k_r$ and two out of the four mass contributions of the perturbation $\delta M=0.1\%,0.5\%$. On the left the results measured by detector $D_1$  and on the right those measured with detector $D_2$.}
\label{fig:mode20}
\end{figure}

\begin{figure}
\centering
\includegraphics[width= 4.25cm]{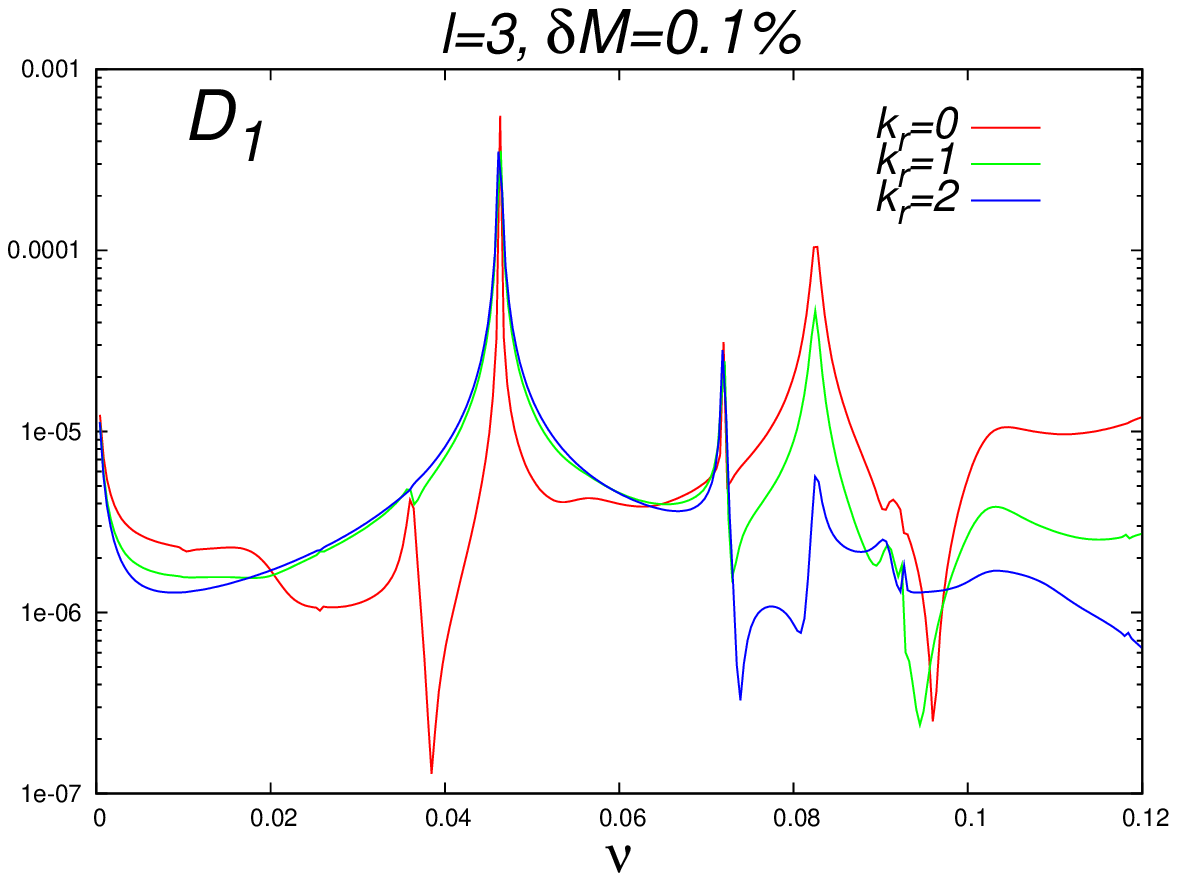}
\includegraphics[width= 4.25cm]{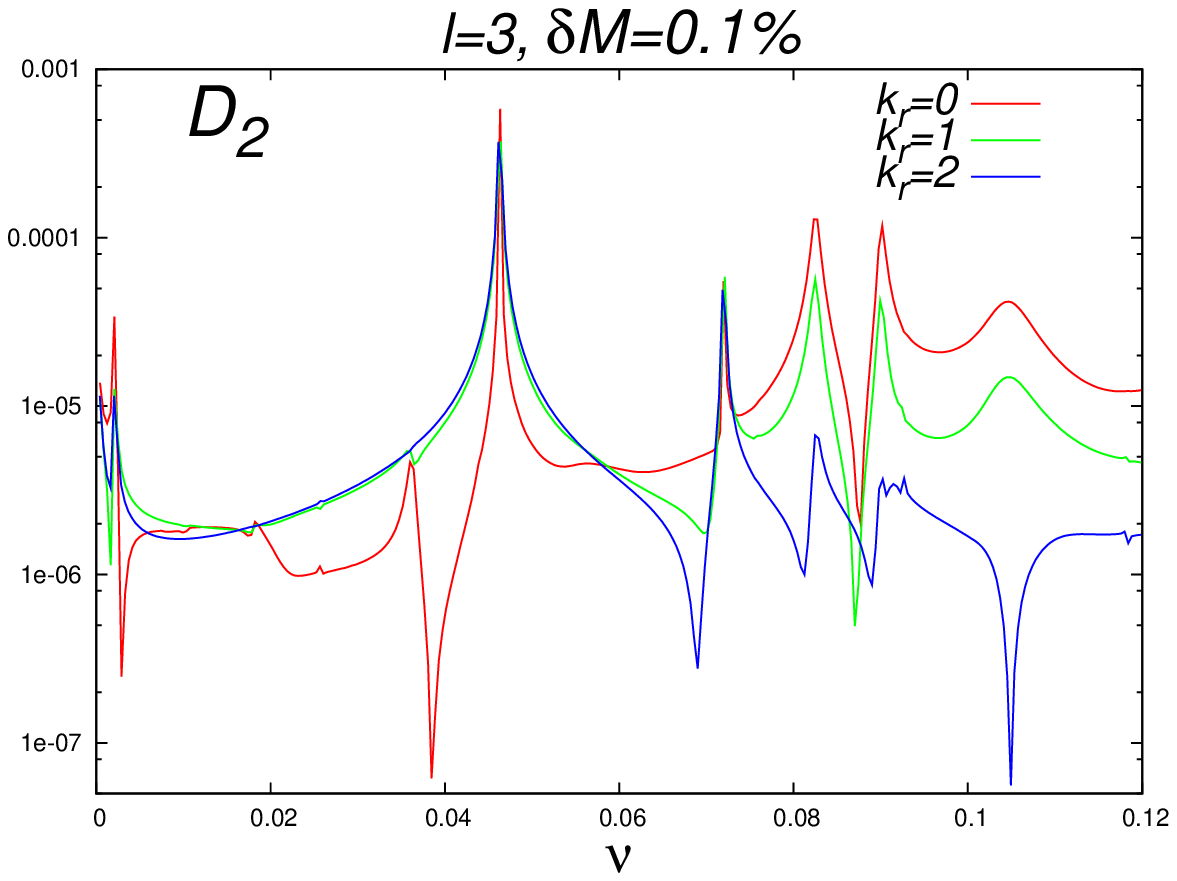}
\includegraphics[width= 4.25cm]{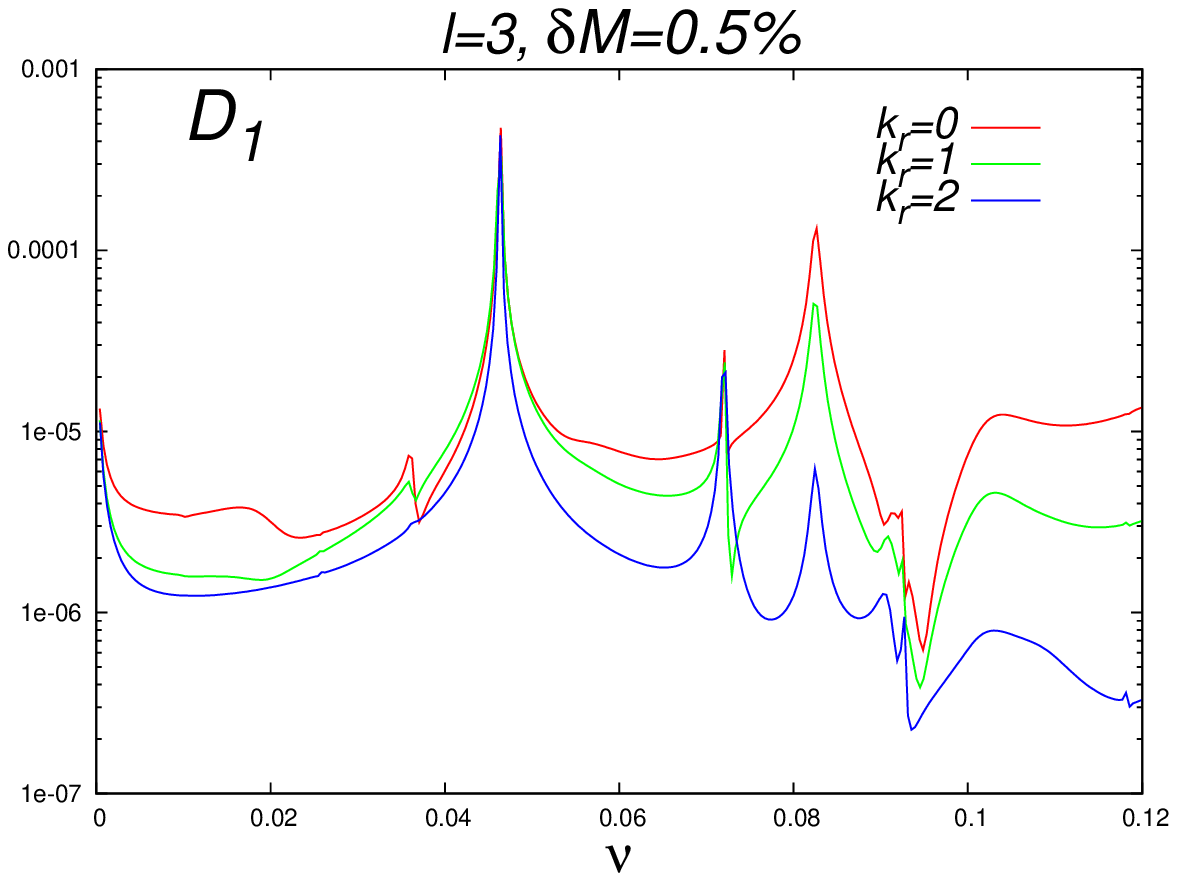}
\includegraphics[width= 4.25cm]{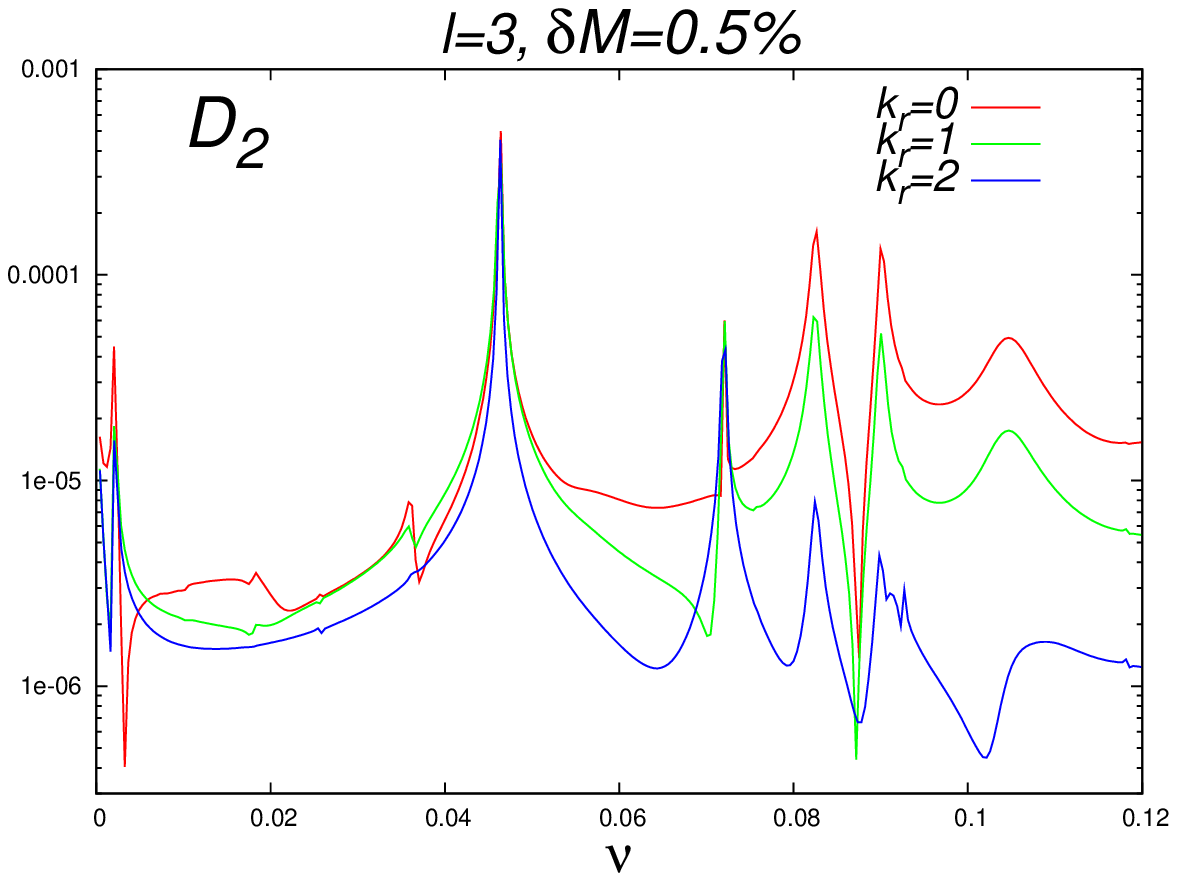}
\caption{Spectrum of the perturbations for the $l=3$ mode for $\sigma_p=1$, $\rho_{max}=20$, the three values of the wave number $k_r$ and two out of the four mass contributions of the perturbation $\delta M=0.1\%,0.5\%$. On the left the results measured by detector $D_1$  and on the right those measured with detector $D_2$.}
\label{fig:mode30}
\end{figure}

\begin{figure}
\centering
\includegraphics[width= 4.25cm]{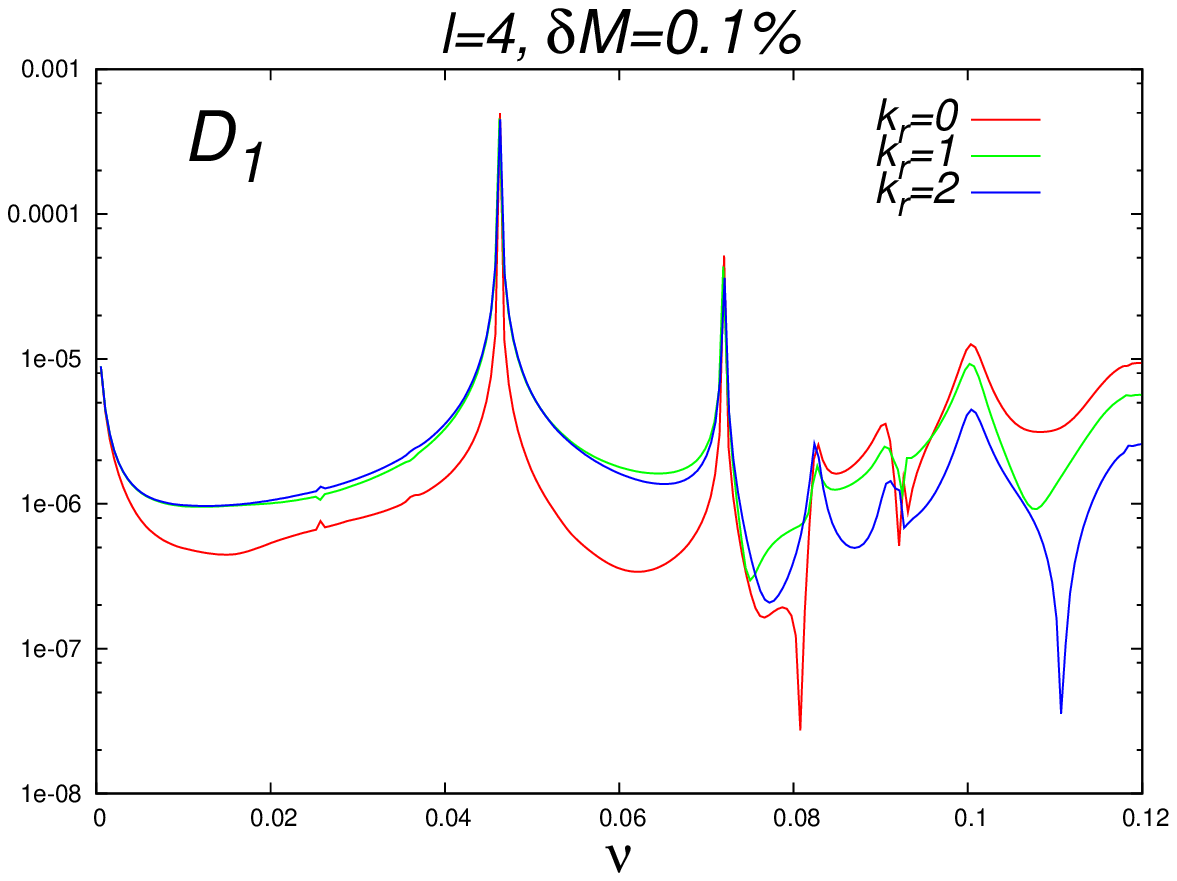}
\includegraphics[width= 4.25cm]{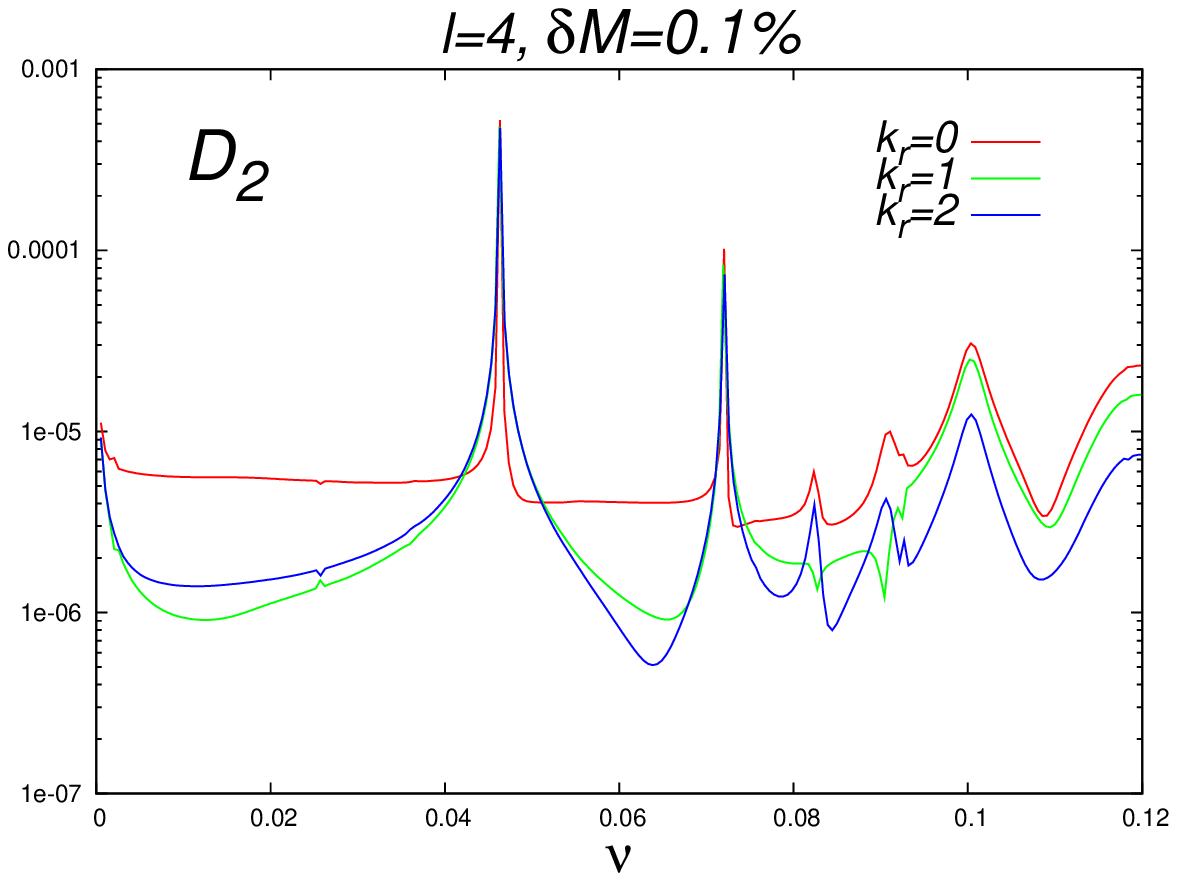}
\includegraphics[width= 4.25cm]{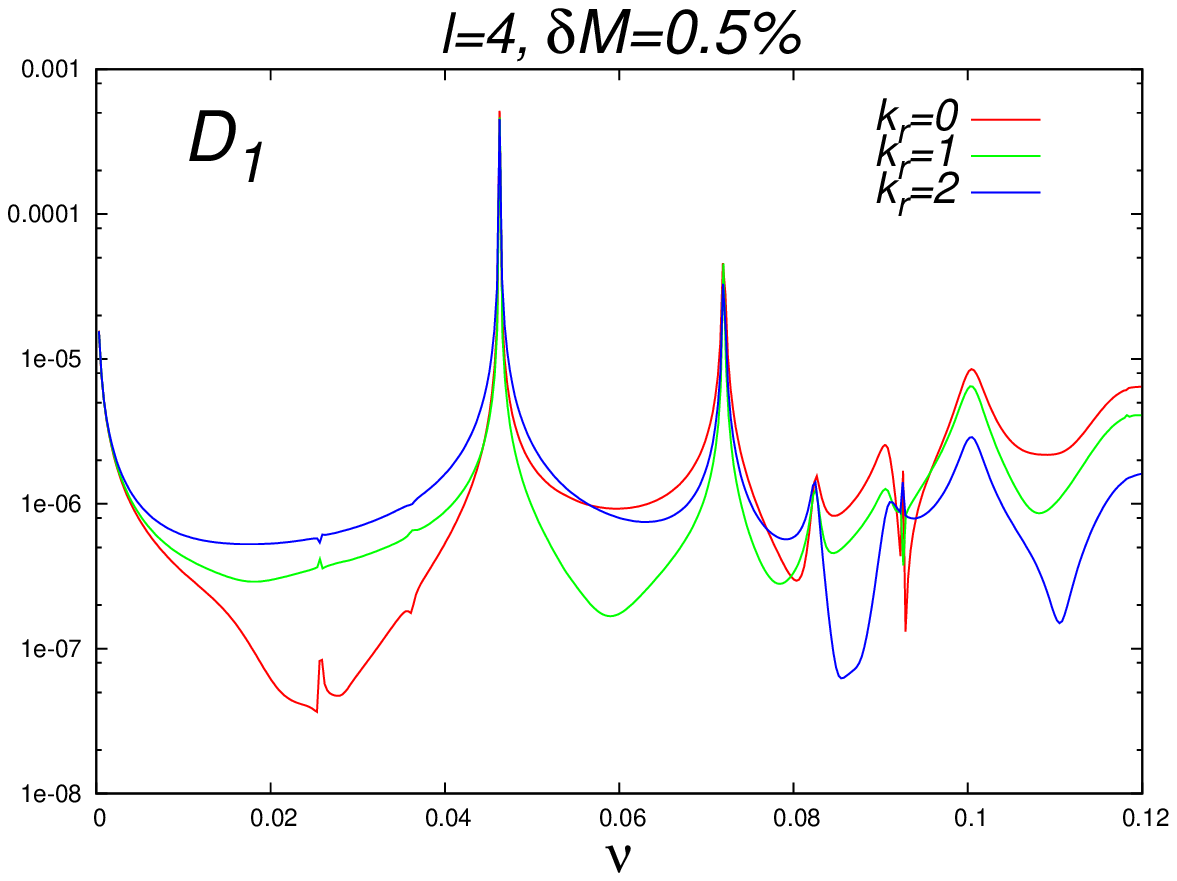}
\includegraphics[width= 4.25cm]{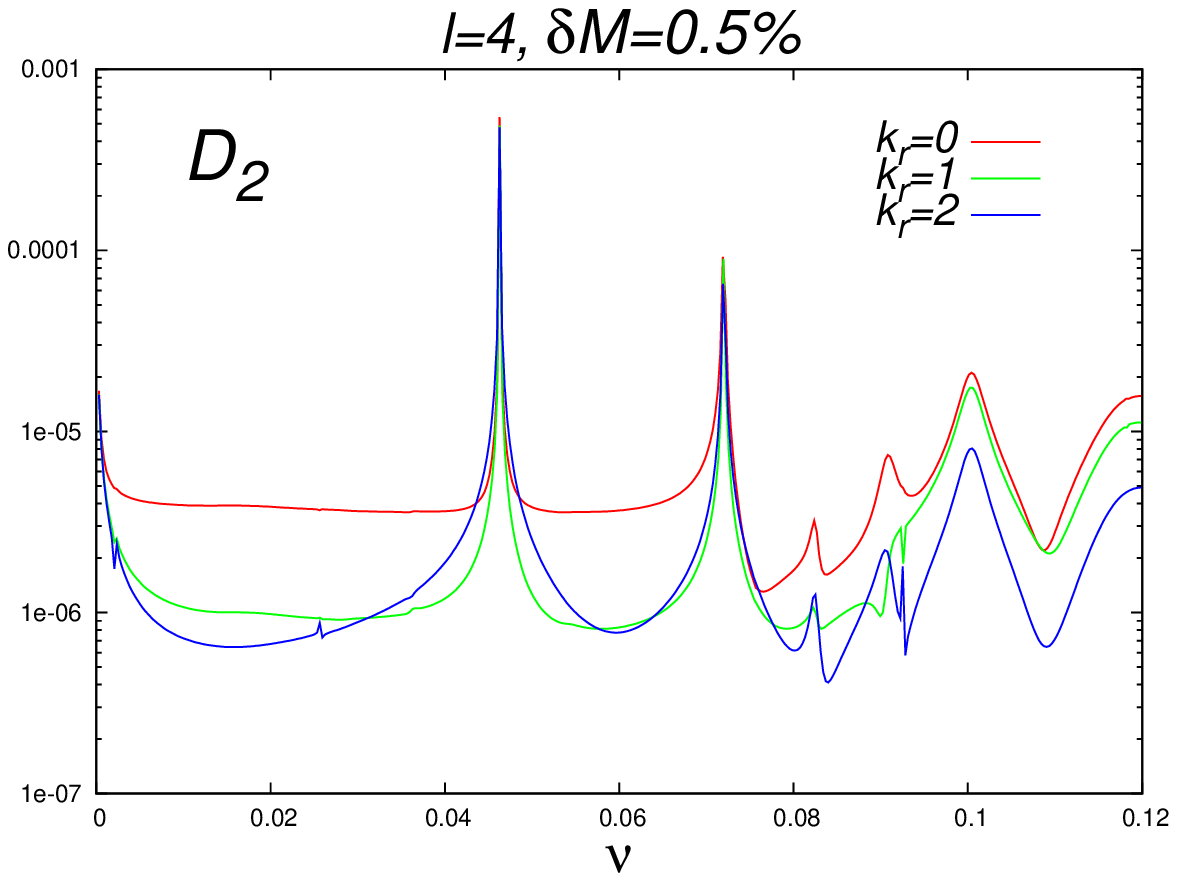}
\caption{Spectrum of the perturbations for the $l=4$ mode for $\sigma_p=1$, $\rho_{max}=20$, the three values of the wave number $k_r$ and two out of the four mass contributions of the perturbation $\delta M=0.1\%,0.5\%$. On the left the results measured by detector $D_1$  and on the right those measured with detector $D_2$.}
\label{fig:mode40}
\end{figure}

\begin{table}
\begin{tabular}{|c|c|c|c|c|}\hline
Mode			&		Detector $D_1$		& Detector $D_2$ 	& $\Delta$	&  	T (Gyr)\\\hline
$l=0$		&		${}^{00}\nu_1 = 0.0462$	&	${}^{00}\nu_1 = 0.0462$	& 0.00004&1.65\\
				&	${}^{00}\nu_2 = 0.0817$	&	${}^{00}\nu_2 = 0.0817$	& 0.0005&  0.93\\\hline
$l=1$		&										&	${}^{10}\nu^{z}_0=0.0019$ &$0.0002$& 37.1\\
			&		${}^{10}\nu^{\varrho}_1 = 0.0462$	&	${}^{10}\nu^{z}_1 = 0.0462$	&$0.0001$& 1.65\\
			&		${}^{10}\nu^{\varrho}_2 = 0.0722$	&		& $ 0.0006$&1.05\\
			&									&	${}^{10}\nu^{z}_2 = 0.0756$ & $ 0.0002$& 1\\\hline
$l=2$	&		${}^{20}\nu^{\varrho}_1=0.0462$		&	${}^{20}\nu^{z}_1=0.0462$	&$0.0001$&1.65\\
			&		${}^{20}\nu^{\varrho}_2=0.0720$		&	${}^{20}\nu^{z}_2=0.0721$ 	&$0.0002$ &1.05\\
			&		${}^{20}\nu^{\varrho}_3=0.0909$		&	${}^{20}\nu^{z}_3=0.0908$ 	& $0.0003$&0.84 \\\hline
$l=3$	&				&	${}^{30}\nu^{z}_0=0.0021$	&$ 0.0001$&37.1\\
			&		${}^{30}\nu^{\varrho}_1 = 0.04635$	&	${}^{30}\nu^{z}_1 = 0.0463$	&$0.0011$& 1.65 \\
			&		${}^{30}\nu^{\varrho}_2 = 0.0720$	&	${}^{30}\nu^{z}_2 = 0.0720$	& $ 0.0008$&1.05\\
			&		${}^{30}\nu^{\varrho}_2 = 0.0912$	&	${}^{30}\nu^{z}_2 = 0.0912$	& $ 0.0009$& 0.84\\\hline
$l=4$	&		${}^{40}\nu^{\varrho}_1=0.0462$		&	${}^{40}\nu^{z}_1=0.0462$	&$0.0005$&1.65\\
			&		${}^{40}\nu^{\varrho}_2=0.0719$		&	${}^{40}\nu^{z}_2=0.0719$ 	&$0.0002$ &1.05\\\hline
\end{tabular}
\label{tab:spectrum}
\caption{Frequencies measured along the radial direction with detector $D_1$ and along the $z-$axis with detector $D_2$. $\Delta$ is the standard deviation of the frequency for each peak, calculated with the results of all the combinations of parameters for the associated frequency of each mode. At the right column the period of oscillation of each mode assuming $m=2.5\times10^{-22}eV$ and core radius $r_c=1$kpc.}
\end{table}

\subsection{Perturbations with modes $l=2$, $l=3$ and $l=4$}

This time we use  (\ref{eq:initialwfnonspherical}) with $l=2,3,4$. Again, the density is measured at $D_1$ and $D_2$ and the FT of the resulting time-series are shown in Fig. \ref{fig:mode20} for $l=2$, in Fig. \ref{fig:mode30} for $l=3$ and in Fig. \ref{fig:mode40} for $l=4$. These examples correspond to the particular case of $\sigma_p=1$, $\rho_{max}=20$, the three values of the wave number $k_r$ and two out of the four mass contributions of the perturbation $\delta M=0.1\%,0.5\%$. For all the wave numbers $\sigma_p$, $\varrho_{max}$, $k_r$ and $\delta M$ there are various frequencies that are triggered and are shown in the Table.

The frequencies in Table I were constructed as follows. For each value of $l$ a batch of runs was performed with all combinations of $k_r,~\sigma_p,~\varrho_{max},~r_p$ and $\delta M$, and for each run the FT was calculated. For each of these FTs the peak frequencies were found and averaged among the results in the batch. The number $\Delta$ is only the standard deviation calculated with each peak frequency within the batch.

\section{Further discussion}
\label{sec:final}

\subsection{Spectroscopy and discussion} 

Collecting the information obtained, the dominant peak frequencies are summarized in Table I. The first interesting frequency is that of the spherical fundamental mode, which is excited by all types of perturbation with frequency $\sim0.0462$.  The perturbation with $l=3$ triggers this mode with an average value slightly off-set from the average value of the peak frequencies, but with contained within the uncertainty box.

The second most excited mode is that with frequency $\sim 0.072$, triggered by all perturbations except $l=0$. In the case $l=1$ it is measured only along the radial direction and along the two axes when applying perturbations with $l=2,3,4$. A specific property of the mode with frequency 0.072, excited by perturbations with $l=2,3,4$ is that it is the dominant peak when the perturbation corresponds to $l=2$ as seen in Fig. \ref{fig:mode20}. This dominance happens not only for the specific parameters used to produce this figure, but also in all the runs carried out for $l=2$, with the various combinations of numerical and perturbation parameters. This is perhaps an important mode that can be triggered by the dynamics of matter, dark or luminous, with an $l=2$ distribution, and eventually usable to characterize cores.

A third and very interesting mode is the one triggered by $l=1$ and $l=3$ perturbations, which has low frequency $\sim0.002$. This mode is measured along the $z$-axis. It has a frequency similar to that found in the density of  configurations resulting from head-on mergers \cite{AvilopGuzman2018b}. The  distribution of matter in  head-on mergers is consistent with the $l=1,m=0$ perturbation mode used here.
This mode becomes  important since head-on mergers result in a single blob whose density oscillates with amplitudes varying by 2 orders of magnitude. A catastrophic result of this scenario is that perhaps luminous matter could not survive such mergers as shown in \cite{GonzalezGuzman2016}, an effect that  could  serve to determine some predictions of this dark matter model or restrict in its viability.

A fourth frequency mode with frequency $\sim0.091$ is found for $l=2$ and $l=3$ perturbations. This is the highest frequency peak described here. The reason for not exploring higher frequency peaks is that the resolution in the frequency domain becomes poor as frequency grows. Finally, there is a mode observed only for the spherical perturbation with frequency $\sim 0.0817$, discovered during the analysis of stability in \cite{BernalGuzman2006b}.

\subsection{Translation to physical units}

A useful method of unit translation is to look at the physical value of the equilibrium configuration eigenvalue $\gamma=0.69223$, for the workhorse configuration with central  density equals to one \cite{GuzmanUrena2004}. This frequency corresponds to $\nu_e=\gamma/(2\pi)\simeq 0.11017$ in code units. In order to estimate the period associated with this frequency in physical units, one fixes the mass of the boson and the radius of the configuration as done in \cite{Mocz2017}, where the period corresponding to $\gamma$ or $\nu_e$  is given by $T_e=6.9\times 10^8 \left( \frac{2.5\times 10^{-22}eV}{m}\right)\left(\frac{r_c}{1kpc}\right)^2$yr, where $r_c$ is the core radius that the density profile of an equilibrium configuration is fitted with, and $m$ is the boson mass. Using this  formula with (for example) $r_c=1$kpc and $m=2.5\times10^{-22}$eV, $T_e=.69$Gyr,  with the translation of $T_e$ into physical units, it is now possible to translate the period of the oscillations triggered by the various perturbation modes, that we show in the right column of the Table. With these values of core radius and boson mass the time scales are of the order of Gyr.  If smaller structures with $r_c=0.1$kpc are considered, the period of each mode will be two orders of magnitude smaller than those in  Table I, which can have an effect on luminous matter on shorter time scales and thus impose restrictions on or make predictions about the model.

Finally a comment on the possible use of the spectrum of oscillations. The frequencies found here are the response to various types of perturbations. The implications of core oscillations are being investigated, for instance in relation to  the possible formation of old star clusters near galactic cores. These analyses impose constraints to the mass of bosonic dark matter and represent state of the art studies \cite{Marsh2018}. So far those investigations are restricted to the diffusion regime, which is based on the consideration of a whole range of core oscillation frequencies. Now, most of the frequencies in our present analysis, except by the very low frequency ${}^{10}\nu_{0}^{z}$ mode, differ at most by only a factor of two, therefore the implications of analyses assuming a whole {\it range} of frequencies, would at the moment be unable to distinguish  one mode from another mode in the spectrum presented here. However, with the results found in this paper, we expect that  observable consequences of core oscillations can start being analyzed considering specific modes and more accurate constraints can be possibly found.


\section*{Acknowledgments}
This research is supported by grants CIC-UMSNH-4.9, CONACyT 258726 and CONACyT 106466.


\end{document}